\documentclass[final,5p,times,twocolumn]{elsarticle}
\usepackage{lineno}
\usepackage{graphicx}
\biboptions{numbers,sort&compress} %compress numbers. citation range.
\usepackage[integrals]{wasysym} %integral
\modulolinenumbers[5]

\usepackage{amssymb, hyperref}
\usepackage{amsmath}
\usepackage{lineno}

\begin{document}

\begin{frontmatter}

\title{Modulation transfer spectroscopy of $^7$Li D1 line}

%% Group authors per affiliation:
\author[JIHT]{Leonid Khalutornykh}
\author[JIHT,HSE]{Sergey Saakyan\corref{cor}}
\ead{saakyan@ihed.ras.ru}
\author[JIHT]{Alexander Nazarov}
\author[JIHT]{Boris B. Zelener}

\address[JIHT]{Joint Institute for High Temperatures, Russian Academy of Sciences (JIHT RAS), Izhorskaya St. 13 Bld. 2, Moscow 125412, Russia}
\address[HSE]{National Research University Higher School of Economics (NRU HSE), Myasnitskaya Ulitsa 20, Moscow 101000, Russia}
\cortext[cor]{Corresponding author}

\begin{abstract} We present the first implementation of modulation transfer spectroscopy (MTS) on the D$1$ line of $^7$Li, carried out in a compact heat-pipe vapor cell with cold windows. 
By varying the pump and probe intensities and polarization configurations, we systematically map the MTS error‑signal amplitude, effective linewidth, and slope for the ground‑state crossover resonance $F=1\times2\rightarrow F'=2$. 
We observe Rabi‑induced power broadening and identify optimal conditions for laser frequency stabilization via tightly focused beams with total power below 1~mW.
The lin$\perp$lin polarization configuration yields a sharp, symmetric, high‑contrast error signal with a maximal slope.
Our findings establish the MTS spectrum of the $^7$Li D1 line transitions as a reliable frequency reference for quantum technology and ultracold atom experiments, particularly in scenarios where frequency stabilization must be achieved with low laser optical power.

\end{abstract}


\end{frontmatter}

\section{Introduction}

Frequency-stabilized narrowband laser radiation sources are essential components of modern quantum technologies. 
These sources are widely employed in precision laser spectroscopy and metrology, where compact and automated frequency-lock systems are necessary for transportable quantum sensors~\cite{degen2017quantum}, such as atomic interferometers~\cite{berg2015composite}, gravimeters~\cite{AfanasievJETPLett, AfanasievUFN,osipenko2023offset}, and vapor cell sensors for magnetic~\cite{magnetometr2022} and radio-frequency (RF) fields~\cite{Nature_rydberg_sensors2024, ryabtsev2025quantum}. 
Compact and robust frequency-stabilized lasers are components of lithium-based primary vacuum sensors~\cite{makhalov2016primary,vacuum2024cold}.

Various frequency locking techniques have been developed to achieve high stability and narrow spectral linewidths, including lock-in detection, polarization spectroscopy, saturated absorption spectroscopy (SAS), and Pound-Drever-Hall (PDH) locking~\cite{tutorial2024}. 
In scenarios requiring wide feedback bandwidths, high-frequency (HF) modulation becomes essential. 
However, in applications where unmodulated laser output is necessary, direct modulation of the laser diode current is unsuitable.

Modulation transfer spectroscopy (MTS) is an advantageous alternative. 
The MTS relies on a coherent nonlinear four-wave mixing process, where modulation is transferred from a pump to a probe beam through a nonlinear resonant medium. 
This method offers several benefits, including a zero Doppler background, broad bandwidth~\cite{so2019zeeman, zhe2011modulation}, and reduced sensitivity to intensity and polarization fluctuations of laser radiation~\cite{MTS_Li_sun2016}. 
Furthermore, MTS generates an error signal in the high-frequency domain, mitigating the influence of low-frequency noise~\cite{zhe2011modulation}. 
MTS has been successfully implemented in commercial laser stabilization systems. 
Compact MTS frequency locking modules for rubidium gravimeters~\cite{lee2021compact} and microcontroller-based self-locking systems~\cite{ruksasakchai2022microcontroller} have already been introduced.

The narrow hyperfine splitting of the excited state makes it difficult to resolve D2 line transitions of lithium with common spectroscopic techniques suitable for laser frequency stabilization and limits the observation of coherent effects~\cite{saakyan2023rydbergD1}.
In MTS signals, the cycling transition dominates even when the hf-splitting of the excited state is comparable to the natural linewidth, as observed for the D2 lines of potassium~\cite{mudarikwa2012sub} and lithium~\cite{MTS_Li_sun2016}. 
In contrast, the choice of the D1 line is advantageous for applications that require a resolved hyperfine structure in the excited state, such as gray molasses and lambda-enhanced cooling techniques~\cite{D1lambda_cooling, prudnikov2023deep}, despite that it involves noncycling transitions.
Recent studies have extended MTS to noncycling transitions of rubidium~\cite{MTS_JQSRT2024, khan2025neighboring, zhao2025synchronous} and potassium~\cite{innes2024modulation}.
MTS resonances on open (non-cycling) transitions are typically an order of magnitude weaker than those on cycling transitions because hyperfine optical pumping depletes population from the addressed ground state~\cite{zhe2011modulation}.
Unlike rubidium and cesium, lithium and potassium exhibit ground-state hyperfine splittings that are smaller than their Doppler-broadened linewidths, which enables the observation of ground-state crossover resonances in pump–probe spectroscopy schemes. In this situation atoms belonging to a single velocity group (with the same absolute value of speed) fulfill both resonance conditions required for the crossover transition, whereas the on-resonant signals originate only from atoms with essentially zero longitudinal velocity. Because the pump and probe address both hyperfine states simultaneously optical pumping into non-absorbing ground levels is suppressed, so the crossover transition behaves as effectively closed.

In this work, we present a detailed investigation of modulation transfer spectroscopy (MTS) on the D1 line of $^7$Li using a high-temperature glass cell with cold windows.
To the best of our knowledge, the MTS spectrum of the $^7$Li D1 line has not previously been reported.  
We explore the characteristics of the MTS signal with the goal of developing a simple and robust setup for laser frequency stabilization.  
By tightly focusing the laser beams into the vapor cell, we obtain MTS signals suitable for laser locking while using less than 1~mW of total pump and probe beam power.
Various polarization configurations of the pump and probe beams are systematically investigated to determine the configuration that yields the strongest and steepest MTS signal for frequency stabilization. 
Special attention is paid to the crossover resonance $2^2\text{S}_{1/2}(\text{F}=1\times2)\rightarrow2^2\text{P}_{1/2}(\text{F}'=2)$, for which we measure the amplitude, the spectral width, and the slope over different pump and probe intensities to evaluate the optimal conditions for robust laser locking.

\section{Theory}
A comprehensive theoretical description of MTS spectra can be found in works~\cite{schenzle1982phase,MTS_JQSRT2024,innes2024modulation,zhao2025synchronous}, including detailed calculations for noncycling transitions for the D1 line of rubidium~\cite{MTS_JQSRT2024} and potassium~\cite{innes2024modulation}. 
The level structure of potassium is very similar to the level structure of lithium.
Here, we briefly outline the main concepts of MTS based on third-order perturbation theory for simple two-level atoms~\cite{shirley1982modulation} and derive simplified expressions to approximate our experimental results.

The electrical field at the intersection of the pump and probe beams when the pump beam is phase-modulated can be expressed as
\begin{multline}
    E=E_{\text{pump}}\sum_{n=0}^{\infty}{J_n(\delta)\exp{(-ikz-i\omega t+in\Omega)}}+\\
    +E_{\text{pr}}\exp{(ikz-i\omega t)}+c.c.
\end{multline}
where $E_{\text{pump}}$ is the amplitude of the original pump field, $E_{\text{pr}}$ is the amplitude of the probe field, $k$ is the laser wavenumber, $\omega$ is the laser carrier frequency, $\Omega$ is the modulation frequency, and $J_n(\delta)$ is the $n$-th order Bessel function whose argument $\delta$ is the modulation index.

According to~\cite{shirley1982modulation} the signal pattern that can be obtained by a photodetector is
\begin{multline}\label{eq2}
    S(\Delta)\propto \sum_{n=-\infty }^{\infty}\dfrac{J_{n}(\delta)\, J_{n-1}(\delta)}{\sqrt{\gamma^2+\Omega^2}}\, \Bigr( (L_{\frac{n+1}{2}}+L_{\frac{n-2}{2}})\cos(\Omega t+\phi) +\\ +(D_{\frac{n+1}{2}}-D_{\frac{n-2}{2}})\sin(\Omega t +\phi)\Bigl),
\end{multline}
where $L_n=(1+x_n^2)^{-1}$, $D_n=x_n\, L_n$, $x_n=2(\Delta-n\Omega)\gamma^{-1}$, $\gamma$ is an effective linewidth incorporating all relevant decoherence mechanisms, $\Delta$ represents the frequency detuning from the line center, and $\phi$ is the relative demodulation phase with respect to the modulation field that is applied to the pump beam. 
In practice, $\phi$ is the sum of the atomic phase shift $\phi_{\mathrm{int}}=\arctan(-\Omega/\gamma)$~\cite{shirley1982modulation} and an externally adjustable electronic phase shift $\phi_{\mathrm{ext}}$ introduced in the demodulation reference path, so that $\phi=\phi_{\mathrm{int}}+\phi_{\mathrm{ext}}$.  
Only the second term is controlled in the experiment.
The equation is simplified if we assume that $\delta<1$, thus, we can consider only two sidebands of the pump field:
\begin{multline}\label{eq3}
    S(\Delta) = C \dfrac{J_0(\delta)\, J_1(\delta)}{\sqrt{\gamma^2+\Omega^2}}\,  \Bigr( (L_{1}+L_{-1/2}-L_{1/2}-L_{-1})\cos(\Omega t+\phi) +\\ +(D_1-D_{1/2}-D_{-1/2}+D_{-1})\sin(\Omega t +\phi)\Bigl)
\end{multline}
where $C$ is a constant amplitude factor.
It is important to note that according to the equation, the amplitude is highly dependent on the phase and modulation index of the signal.
The Eq.~\ref{eq2} are derived for weak intensities.
For higher intensities, we assume that saturation only modifies an effective linewidth and an overall scale, and that the signal remains a weighted sum of $L$ and $D$ components.
Accordingly, the experimental spectra are fitted with the sum Lorentzian and dispersive functions, treating the effective linewidth $\gamma$, $C$, and $\phi$ as free parameters.

\section{Experimental setup and methods}

Spectroscopic measurements of the D1 line in $^7$Li atoms were performed using a modified heat-pipe oven design~\cite{vidal1969heat, heatpipe1973}. 
The gas cell configuration consists of a 55~cm long quartz tube (2.5~cm diameter) containing a 20~cm long molybdenum foil sleeve that houses a piece of metallic lithium.
The main difference from the standard design is that the active cooling systems for its windows were eliminated, resulting in a more compact and simple setup.

The working region was resistively heated to 435~$^\circ$C using a nichrome wire coil, whereas passive thermal dissipation maintained the quartz windows at room temperature. 
This thermal gradient, together with the buffer gas (argon) being pumped into the cell volume at a pressure of 0.01~Torr, prevents the lithium metal film from depositing on the windows.
Temperature monitoring was achieved through a K-type thermocouple positioned between the external quartz surface and the heating elements.
The heating element has a noninductive spiral winding geometry to minimize magnetic field generation. 
It is housed in a fireclay insulation system with additional aluminum foil wrap to prevent heat loss.
Spectral analysis confirmed that residual stray magnetic fields in the interaction region exerted negligible influence on both SAS and MTS signals~\cite{MTS_Li_sun2016}.

Temperature regulation ($\pm1$~$^\circ$C stability) was implemented through a Wi-Fi enabled ESP8266 microcontroller (NodeMCU v3) interfaced with a programmable DC power supply via the RS-232 protocol. 
A web interface hosted on the microcontroller is used to adjust the temperature.

\begin{figure}[t]
    \centering
    \includegraphics[width=7 cm]{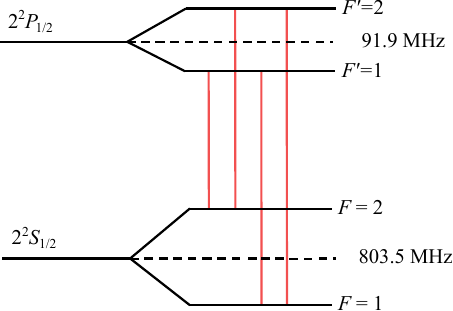}
    \caption{The level structure of the D1 line in $^7$Li. The diagram shows the $2^2S_{1/2} \rightarrow 2^2P_{1/2}$ transition, including the hyperfine splitting of both the ground and excited states. The red vertical lines represent the allowed electric dipole transitions between the hyperfine levels. The frequency intervals between hyperfine states are indicated on the right.}

    \label{fig1}
\end{figure}

Fig.~\ref{fig1} shows the energy level diagram of the $^7$Li D1 transitions.
Spectroscopic measurements were performed using an external cavity diode laser coupled with a tapered amplifier (Toptica DLC TA Pro). 
The laser operates at 670 nm with a maximum output power of 500 mW.
The laser frequency was scanned across the $^7$Li D1 hyperfine transitions at a rate of 50~Hz.

\begin{figure}[t]
    \includegraphics[width=\linewidth]{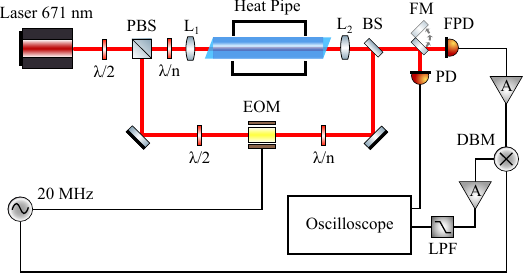}
    \caption{Schematic of the experimental setup.
    A laser beam is split into probe and pump beams using a half-wave plate ($\lambda/2$) with a polarizing beam splitter (PBS). 
    The probe beam traverses the lithium gas cell and is directed to either a photodiode (PD) or a fast photodiode (FPD). 
    A flip mirror (FM) swithces between the MTS and SAS detection setups. 
    Lenses L$_1$ (35~cm) and L$_2$ (40~cm) focus the beams into the interaction region. 
    The counter-propagating pump beam is phase-modulated at $\Omega=20$~MHz by an electro-optic modulator (EOM). 
    The half-wave plate ($\lambda/2$) before the EOM ensures fine alignment of the pump beam polarization with the EOM axis. 
    The wave plates labeled $\lambda/n$ ($n=2,4$) configure different polarization states for pump and probe beams. 
    A nonpolarizing beam splitter (BS) combines the counter-propagating beams. 
    The FPD signal is amplified (A), processed through a double-balanced mixer (DBM), filtered by a 30~kHz low-pass filter (LPF), and recorded by an oscilloscope.}
    \label{fig2}
\end{figure}

A schematic of the experimental setup for MTS and SAS detection for different polarization configurations of pump and probe beams is shown in Fig.~\ref{fig2}.
The laser output is split into two orthogonally polarized beams via a polarizing beam-splitter cube (PBS). 
A fast photodetector (FPD, model ET-2030A from EOT) detects a probe beam passing through a PBS cube and a cell containing lithium vapor.
The counter-propagating pump beam is phase-modulated at 20~MHz by an electro-optical modulator (EOM, model EO-AM-NR-C1 from Thorlabs) with a modulation index of $\approx0.28$, corresponding to a sideband intensity of approximately 2\% of the total pump beam intensity. 
Lenses were employed to minimize the total power required for MTS and laser stabilization.
The beam intensities were determined by measuring the beam waist and Rayleigh range using a CCD camera. 
The geometric mean waist of the slightly astigmatic pump (probe) beam was found to be $w_{0}=84(115)~\mu$m, with a corresponding Rayleigh range of $z_{\text{R}}=3.1(8.1)$~cm.
Following beam size characterization, the gas cell was positioned and centered at the beam waist.

The probe beam signal is detected by a fast photodetector (FPD), amplified using a high-frequency amplifier (model ZFL-1000LN+, Mini-Circuits), and subsequently demodulated by a double-balanced mixer (DBM, model ZAD-1+ from Mini-Circuits). 
To drive the EOM, we use one output channel of a two-channel waveform generator (model DG4102, Rigol) without additional amplification or the use of a resonant RLC circuit~\cite{mccarron2008modulation}. 
The second output of the generator serves as the local oscillator for the DBM and enables phase adjustment of the MTS signal. 
Both channels of the signal generator are internally phase-locked to a common reference oscillator.
The mixer output is amplified and passed through a 30 kHz low-pass filter. 
A flip mirror (FM) mount is used to switch between MTS and saturated absorption spectroscopy (SAS) detection schemes. 
Both MTS and SAS signals are recorded and averaged using a digital oscilloscope.

\begin{figure}[t]
    \includegraphics[width=\linewidth]{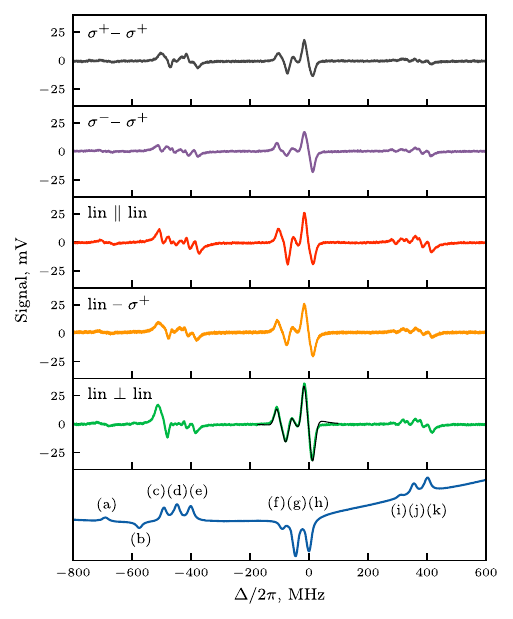}
    \caption{
    Modulation transfer spectra for various polarization configurations of the pump and probe beams.  
    The black curve in the lin$\perp$lin panel represents the best fit by Eq.~(\ref{eq3}).  
    The bottom panel shows the SAS spectrum with labeled D2 line transitions of $^6$Li~(a–b) and D1-line transitions of $^7$Li~(c–k).  
    The symbol $\times$ denotes a crossover resonance, whereas $F$ and $F'$ refer to the hyperfine levels of the ground and excited states, respectively.  
    The labeled features correspond to the following transitions:  
    (a)~$F = 3/2 \rightarrow F' = 1/2$, $3/2$, $5/2$;  
    (b)~$F = 3/2 \times 1/2 \rightarrow F' = 1/2$, $3/2$, $5/2$;
    (c)~$F = 2 \rightarrow F' = 1$;
    (d)~$F = 2 \rightarrow F' = 1 \times 2$;
    (e)~$F = 2 \rightarrow F' = 2$;
    (f)~$F = 1 \times 2 \rightarrow F' = 1$;
    (g)~$F = 1 \times 2 \rightarrow F' = 1 \times 2$;
    (h)~$F = 1 \times 2 \rightarrow F' = 2$;
    (i)~$F = 1 \rightarrow F' = 1$;
    (j)~$F = 1 \rightarrow F' = 1 \times 2$;
    (k)~$F = 1 \rightarrow F' = 2$.
    The resonance labeled~(g) corresponds to an $N$-scheme crossover, which does not involve any shared energy levels between the contributing transitions~\cite{scotto2015four}.  
    The transition $F = 3/2 \rightarrow F' = 1/2, 3/2$ of the $^6$Li D2 line, which lies between features~(c) and~(d), is not visible owing to its weak signal strength.
    }
    \label{fig3}
\end{figure}

Fig.~\ref{fig3} presents the SAS spectrum alongside the MTS spectra obtained under various polarization configurations of the pump and probe beams.
These configurations were adjusted using combinations of quarter-wave and half-wave plates (denoted as $\lambda/n$, with $n = 4$ and $2$, respectively, in Fig.~\ref{fig2}). 
All subsequent measurements were performed using the $F=1\times2 \rightarrow F'=2$ crossover resonance of the $^7$Li D1 line, which has the highest amplitude among the observed dispersion-like curves. 
The linear orthogonal (lin$\perp$lin) polarization configuration was selected because of its superior peak-to-peak amplitude compared to those of the other polarization arrangements.
Under the lin$\parallel$lin~configuration, the $\pi$-polarized pump and probe address only $\Delta m = 0$ transitions.
Continuous optical pumping therefore drives population into the dark ground-state
sublevel $|F = 2,\,m_F = 0\rangle$, weakening the crossover feature.
Employing orthogonal polarizations (lin$\perp$lin) introduces $\sigma^{\pm}$ components into the probe beam, which address the $\Delta m = \pm 1$ transitions and continuously redistribute population among the Zeeman sublevels, preventing accumulation in dark states and producing a noticeable enhancement of the crossover amplitude.
In the $\sigma^{+}-\sigma^{-}$ and $\sigma^{+}-\sigma^{+}$ beam configurations the Zeeman sub-levels $\lvert F = 2,\, m_{F} = \pm F \rangle$ are optically pumped into non-interacting (dark) states. The lin–$\sigma^{+}$ configuration is similar to the lin$\perp$lin configuration, except that the pump drives $\Delta m = +1$ transitions rather than $\Delta m = 0$ in the perpendicular case. 
The difference in signal amplitude between these two configurations can be attributed to their distinct excitation pathways, transition probabilities, and resulting population distributions.

Full density-matrix calculations for the potassium D1 line~\cite{innes2024modulation} show that the perpendicular linear configuration offers the strongest MTS crossover signal. 
The exact amplitude ratio between any two polarization configurations depends on how population is restored relative to the optical-pumping rate. 
Restoring is governed by several mechanisms, including the transit time of ground-state atoms through the beam, collisions, buffer-gas pressure in the cell, and residual stray field environment.

Orthogonal polarizations are particularly advantageous since they mitigate interference patterns and noise introduced by unwanted reflections and standing waves within optical components. While this etalon effect is significantly suppressed the demodulated MTS signals and the reduced residual amplitude modulation (RAM) of the background signal become much cleaner.

We extract the parameters of the MTS spectrum by fitting the crossover resonances with the following function:
\begin{equation}\label{eq4fit}
    \sum_{i=1}^{3} S_i(\Delta + \Delta_{0i}),
\end{equation}
where $C_i$, $\gamma_i$, $\phi_i$, and $\Delta_{0i}$ are treated as free parameters. 
Here, $\Delta_{0i}$ is the detuning offset required to take into account the relative frequency shift of the transitions. This procedure follows Refs.~\cite{eble_Ca,mccarron2008modulation,preuschoff2018optimization}, which all demonstrate that the same empirical model can describe data equally well, even at high total intensities. These crossover features correspond to the transitions $F=1\times2 \rightarrow F'=1$, $F=1\times2 \rightarrow F'=1\times2$, and $F=1\times2 \rightarrow F'=2$ of the D1 line of $^7$Li.
They are labeled as~(f),~(g), and~(h), respectively, in the lower panel of Fig.~\ref{fig3}.

A modulation frequency of $\Omega = 20$~MHz was selected experimentally, as it yields a higher peak-to-peak amplitude of the MTS signal. 
In general, it is recommended to choose a modulation frequency close to $0.35\gamma$~\cite{riehle2006frequency}, where $\gamma$ is the effective linewidth. 
Under our experimental conditions, $\gamma$ includes all relevant decoherence mechanisms, such as Rabi broadening and transit-time broadening~\cite{shirley1982modulation}.
For lithium at a temperature of 435~$^\circ$C, the transit-time broadening is estimated to be less than 2~MHz.

\begin{figure}[t]
    \includegraphics[width=\linewidth]{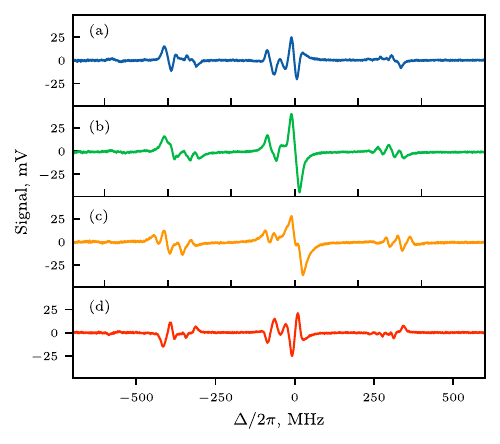}
    \caption{MTS spectra recorded at different phase settings between the modulation signal and the reference oscillator.
    (a) Phase offset of $-60^\circ$ from optimal;
    (b) optimal phase setting, that yields a symmetric, high-contrast error signal suitable for frequency stabilization;
    (c) phase offset of $+60^\circ$ from optimal; (d) phase offset of $+120^\circ$ from optimal.}
    \label{fig4}
\end{figure}

By varying the external phase shift $\phi_{\text{ext}}$ we can select
any linear combination of the absorptive and dispersive line-shapes,
thereby modifying both the symmetry and the amplitude of the demodulated
signal.
Phase adjustment is crucial to obtain a signal suitable for frequency stabilization.
Fig.~\ref{fig4}(b) shows the signal shape considered optimal for this purpose. 
The signal can be optimized by tuning the phase $\phi_\text{ext}$ to achieve a symmetric error signal with maximized amplitude for the ground-state crossover transition $F=1\times2 \rightarrow F'=2$.

\section{Results and discussion}

\begin{figure}[t]
    \includegraphics[width=\linewidth]{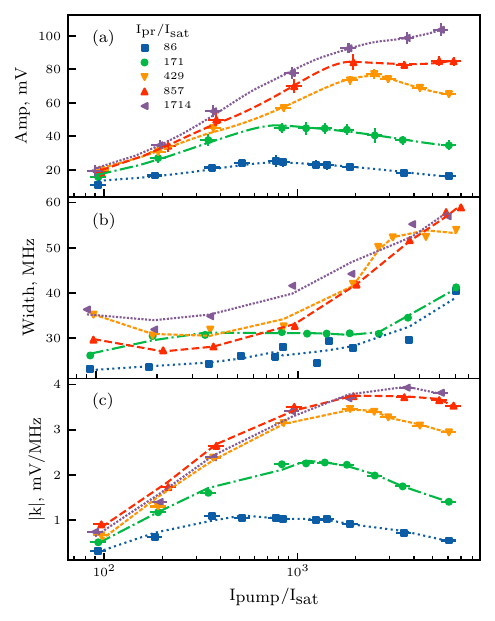}
    \caption{Dependence of the peak-to-peak amplitude~(a), effective linewidth~(b), and absolute slope coefficient~(c) of the error signal on the normalized pump-beam intensity $I_{\text{pump}}/I_{\text{sat}}$. 
    Each curve corresponds to a different probe-beam intensity $I_{\text{pr}}/I_{\text{sat}}$, as indicated in the legend.}
    \label{fig5}
\end{figure}

Fig.~\ref{fig5} shows how the peak-to-peak amplitude, effective linewidth, and absolute value of the slope coefficient of the selected crossover transition $F=1\times2 \rightarrow F'=2$ vary with pump-beam intensity for different probe-beam intensities. 
The effective linewidth was extracted by fitting each measurement of the ground-state crossover resonances with the Eq.~(\ref{eq4fit}), where $\gamma_i$, $\phi_i$, $C_i$, and $\Delta_{0i}$ are treated as free parameters. 
As pump intensity increases, the effective linewidth rises monotonically. 
This increase is in good agreement with Rabi (power) broadening, even at intensities beyond the weak intensity model.

All intensity values refer to the peak intensity of Gaussian beams and are normalized to $I_{\mathrm{sat}} = 2.54$~mW/cm$^2$, which corresponds to the saturation intensity of the cycling transition of the D2 line. 
In this study, the laser beams are tightly focused within the vapor cell, resulting in a Rayleigh range shorter than the effective interaction length inside the heat-pipe. 
The intensity of the modulation sidebands is approximately 2\% of the pump beam intensity.
It is important to note that the actual saturation intensity for noncycling transitions of the D1 line is significantly higher~\cite{bloch1996doppler} than the D2 line value used for normalization.
The total optical power of the pump and probe beams remained below 1~mW during all the experiments.

As shown in Fig.~\ref{fig5}(c), for all selected values of the normalized probe beam intensity $I_{\text{pr}}/I_{\text{sat}}$, there exists an optimal pump beam intensity that maximizes the amplitude of the error signal used for frequency stabilization.  
A higher slope of the demodulated signal is critical for stabilization performance and depends on the beam waist size.
For a comprehensive discussion of how beam size affects frequency stability in laser frequency standards, see Ref.~\cite{lee2023laser}.

\begin{figure}[t]
    \includegraphics[width=\linewidth]{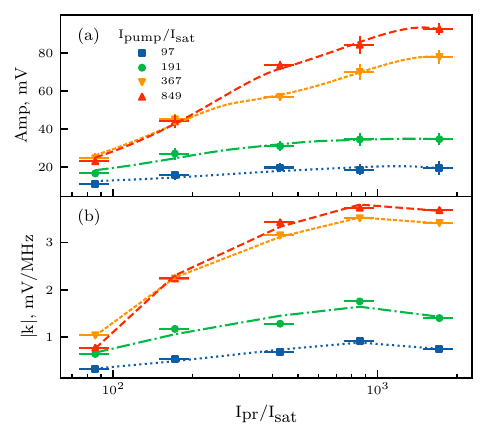}
    \caption{Dependence of the peak-to-peak amplitude~(a) and absolute slope coefficient~(b) of the error signal on the normalized probe-beam intensity $I_{\text{pr}}/I_{\text{sat}}$. 
    Each curve corresponds to a different pump-beam intensity $I_{\text{pump}}/I_{\text{sat}}$, as indicated in the legend.}
    \label{fig6}
\end{figure}

Fig.~\ref{fig6} presents the dependence of both the peak-to-peak amplitude and the absolute slope of the MTS signal on the probe beam intensity.
The effective linewidth shows a similar dependence on the probe intensity as in Fig.~\ref{fig5}(b) and is therefore not displayed. 
The error bars in Fig.~\ref{fig5} and Fig.~\ref{fig6} represent the fitting error. 
The relative error in the beam waist $w_0$ is estimated to be 8\%, primarily due to the beam quality of the diode laser, which in turn contributes to the overall intensity error.
Most of the increase in signal amplitude observed with an increase in the probe intensity is attributed to the enhanced response of the photodetector.
Special care was taken to ensure that the photodetector operated well below its saturation threshold across the entire range of probe intensities.
Increase in the probe beam intensity produces an effect similar to the increase in the pump intensity. The effective linewidth broadens, leading to a corresponding reduction in the slope of the error signal.

As shown in Ref.~\cite{MTS_Li_sun2016}, the crossover peak is more suitable for frequency stabilization due to its improved compensation of RAM in the background signal.  
Additional enhancement is achieved by using orthogonal beam polarizations, which help suppress the RAM by eliminating the etalon effect.
It has been demonstrated in Ref.~\cite{MTS_Li_sun2016} that the lin$\perp$lin configuration yields a stronger MTS signal for ground-state crossover transitions of the D2 line.
In this configuration, all magnetic sublevels of the $F=1$ ground state contribute to the signal, and the absence of forbidden transitions results in a higher amplitude than other polarization schemes do.
Due to the unique level structure of lithium, even the D2 transition is not fully closed (often described as a "weakly closed transition"), as spontaneous decay can populate the other ground-state level because of the unresolved hyperfine structure of the excited state.

\section{Conclusion}

We experimentally demonstrated modulation transfer spectroscopy on the $^7$Li D1 line using a compact quartz vapor cell with cold windows and sub-milliwatt optical powers. The MTS signal associated with the $F=1\times2 \rightarrow F'=2$ crossover resonance was analyzed in terms of its amplitude, width, and slope under varying experimental parameters. Among the tested configurations, the lin$\perp$lin polarization scheme provided the most favorable conditions for laser frequency stabilization, offering a strong and steep error signal. The resonance amplitude and slope increase with pump intensity, while the spectral width broadens accordingly. Our results are consistent with theoretical models of an MTS in a two-level system and demonstrate the feasibility of implementing simple, optically power-efficient, and highly stable laser locking systems for lithium-based experiments. These findings are expected to be valuable for the future development of transportable quantum sensors and laser cooling setups utilizing lithium atoms.

% Description here https://www.elsevier.com/researcher/author/policies-and-guidelines/credit-author-statement
\section*{CRediT authorship contribution statement}
\textbf{Leonid Khalutornykh}: Investigation, Software, Validation, Formal analysis, Visualization, Writing -- original draft, Writing -- review \& editing. 
\textbf{Sergey Saakyan}: Conceptualization, Methodology, Validation, Writing -- original draft, Writing -- review \& editing, Funding acquisition. 
\textbf{Alexander Nazarov}: Validation, Writing -- original draft, Writing -- review \& editing.
\textbf{Boris B. Zelener}: Supervision, Resources, Writing -- review \& editing, Funding acquisition.

\section*{Declaration of Competing Interest}
The authors declare that they have no known competing financial interests or personal relationships that could have appeared to influence the work reported in this paper.

\section*{Data availability}
Data will be made available upon request.

\section*{Acknowledgements}
The author wishes to thank V.A. Sautenkov and D.V. Kazantsev for a very useful discussion. The research has been supported by the Russian Science Foundation Grant No.\,23-72-10031. The development of the vapor cell was supported by the Ministry of Science and Higher Education of the Russian Federation (State Assignment No.\,075-00269-25-00).

\bibliographystyle{elsarticle-num}
\bibliography{refs}

\end{document}